\documentclass[12pt,a4paper]{article}
\usepackage{a4wide,amsmath,cite}

\newcommand{\Tr}{\mathop{\mathrm{Tr}}\nolimits}

\title{Complete analysis of polarization effects
in $e\gamma\to e\gamma$ with REDUCE}
\author{A.~G.~Grozin%
\thanks{Supported in part by the Russian foundation of fundamental research,
code 93--02--14428}\hspace{2mm}%
\thanks{Email: \texttt{A.G.Grozin@inp.nsk.su}}\\
\textit{Budker Institute of Nuclear Physics, Novosibirsk 630090, Russia}}
\date{}

\begin{document}

\maketitle

\begin{abstract}
Formulae for the matrix element squared of the process $e\gamma\to e\gamma$
with all four particles polarized are obtained using REDUCE.
\end{abstract}

Polarization effects in the quantum electrodynamics process
$e\gamma\to e\gamma$ were considered in many works~\cite{BLP,GKPST,T,M,G},
but formulae with all four particles polarized are absent
in the available literature.
These polarization effects are important
for the proposed conversion of high energy electron beams
to photon beams that will allow to obtain $e\gamma$ and $\gamma\gamma$
colliders on the basis of the future linear $e^-e^+$ colliders~\cite{GKPST}.
Here I obtain a complete set of formulae in covariant notations
using REDUCE~\cite{Hearn}.
I follow the method of~\cite{BLP}.

Let the electron mass $m=1$;
it will be restored in the final answers by dimension.
The momenta are $p+k=p'+k'$; $x=s-1$, $y=1-u$; $s=(p+k)^2$, $u=(p-k')^2$.
The matrix element squared is~\cite{BLP}
\begin{equation}
\begin{split}
&\overline{|M|^2} = 64 \pi^2 \alpha^2 \rho'_{\rho\mu} \rho_{\nu\sigma}
\Tr \rho' \frac{Q_{\mu\nu}}{xy} \rho \frac{\overline{Q}_{\rho\sigma}}{xy}\,,\\
&\frac{Q_{\mu\nu}}{xy} = \frac{\gamma_\mu (\hat{p}+\hat{k}+1) \gamma_\nu}{x}
- \frac{\gamma_\nu (\hat{p}-\hat{k'}+1) \gamma_\mu}{y}\,.
\end{split}
\label{M2}
\end{equation}
It is convenient to use the basis
\begin{equation}
n_0 = \frac{K}{v}\,,\quad
n_1 = \frac{q}{v}\,,\quad
n_2 = \frac{v}{2w} P_\bot\,,\quad
n_{3\mu} = - \frac{1}{2vw} \varepsilon_{\mu\alpha\beta\gamma} K_\alpha q_\beta P_\gamma\,,
\label{n}
\end{equation}
where $K=k+k'$, $q=k'-k=p-p'$, $P_\bot=P-\frac{PK}{K^2}K$, $P=p+p'$;
$v=\sqrt{x-y}$, $w=\sqrt{xy-x+y}$;
\begin{equation}
\begin{split}
&p = \frac{1}{v} \left( \frac{x+y}{2} n_0 + \frac{x-y}{2} n_1 + w n_2 \right)\,,\\
&p' = \frac{1}{v} \left( \frac{x+y}{2} n_0 - \frac{x-y}{2} n_1 + w n_2 \right)\,,\\
&k = \frac{v}{2} (n_0-n_1)\,,\quad
k' = \frac{v}{2} (n_0+n_1)\,.
\end{split}
\label{pk}
\end{equation}
The vectors $\vec{n}_1$, $\vec{n}_2$, $\vec{n}_3$ form a right-handed system.

The vectors $n_2$, $n_3$ can be used as polarization vectors
(in the scattering plane and perpendicular to it) of both photons.
For the final photon, the vectors $\vec{n}_2$, $\vec{n}_3$, $\vec{k}\,'$
form a right-handed system (in the $K$ rest frame);
therefore its density matrix is expressed via the Stokes parameters $\xi'_j$
in the standard way: $\rho'_{\mu\nu}=\frac{1}{2}\sum\xi'_j\sigma_{j\mu\nu}$,
where $\xi^{(\prime)}_0=1$ have been formally introduced, and
\begin{equation}
\begin{split}
&\sigma_{0\mu\nu} = n_{2\mu} n_{2\nu} + n_{3\mu} n_{3\nu}\,,\quad
\sigma_{1\mu\nu} = n_{2\mu} n_{3\nu} + n_{3\mu} n_{2\nu}\,,\\
&\sigma_{2\mu\nu} = -i (n_{2\mu} n_{3\nu} - n_{3\mu} n_{2\nu})\,,\quad
\sigma_{3\mu\nu} = n_{2\mu} n_{2\nu} - n_{3\mu} n_{3\nu}\,.
\end{split}
\label{Pauli}
\end{equation}
For the initial photon,
the right-handed system is $\vec{n}_2$, $-\vec{n}_3$, $\vec{k}$.
Therefore
$\rho_{\mu\nu}=\frac{1}{2}\sum_j\delta_j\xi_j\sigma_{j\mu\nu}$,
where $\delta_0=\delta_3=1$, $\delta_1=\delta_2=-1$.

The tensor $Q_{\mu\nu}$ in the $n_2$--$n_3$ plane also can be expanded in
$\sigma$-matrices:
\begin{equation}
\begin{split}
&Q_{\mu\nu} = \sum_k Q_k \sigma_{k\mu\nu}\,,\quad
Q_k = \frac{1}{2} Q_{\mu\nu} \sigma_{k\nu\mu}\,,\\
&\frac{Q_{\mu\nu}}{xy} = - \frac{w}{v}
\left(\frac{\gamma_\mu\gamma_2\gamma_\nu}{x}
- \frac{\gamma_\nu\gamma_2\gamma_\mu}{y}\right)
+ \frac{\gamma_\mu\gamma_0\gamma_\nu-\gamma_\nu\gamma_0\gamma_\mu}{v}
+ \frac{\gamma_\mu\gamma_\nu}{x} - \frac{\gamma_\nu\gamma_\mu}{y}\,.
\end{split}
\label{Q0}
\end{equation}
Using the Dirac equation, we obtain
\begin{equation}
\begin{split}
&Q_0 = x-y\,,\quad
Q_1 = -i (x-y) \gamma_5 \hat{K}/2\,,\\
&Q_2 = -(x-y) \gamma_5\,,\quad
Q_3 = x-y - (x+y) \hat{K}/2\,.
\end{split}
\label{Q}
\end{equation}
The conjugate tensor
$\overline{Q}_{\mu\nu}=\sum_k\delta_k Q_k \sigma_{k\nu\mu}$,
because $\overline{Q}_k=\delta_k Q_k$, $\sigma^*_{k\mu\nu}=\sigma_{k\nu\mu}$.

The initial and final electron density matrices are
$\rho^{(\prime)}=\frac12(\hat{p}^{(\prime)}+1)(1-\gamma_5\hat{a}^{(\prime)})$.
Let's introduce two bases
\begin{equation}
\begin{split}
&e_0 = p\,,\quad
e_1 = p - \frac{2}{x} k\,,\quad
e_2 = \frac{w}{x} (n_0-n_1) + n_2\,,\quad
e_3 = n_3\,;\\
&e'_0 = p'\,,\quad
e'_1 = p' - \frac{2}{x} k'\,,\quad
e'_2 = -\frac{w}{x} (n_0+n_1) - n_2\,,\quad
e'_3 = n_3\,.
\end{split}
\label{e}
\end{equation}
Then $a^{(\prime)}=\sum_i\zeta^{(\prime)}_i e^{(\prime)}_i$,
where $\zeta_1$ is the longitudinal polarization,
$\zeta_2$ is the transverse polarization in the scattering plane,
and $\zeta_3$ is the transverse polarization perpendicular to this plane.
Introducing formally $\zeta^{(\prime)}_0=1$,
we have $\rho^{(\prime)}=\frac{1}{2}\sum_i\zeta^{(\prime)}_i\rho^{(\prime)}_i$,
where $\rho^{(\prime)}_0=\hat{p}^{(\prime)}+1$,
$\rho^{(\prime)}_i=-\rho^{(\prime)}_0\gamma_5\hat{e}^{(\prime)}_i$.

Finally, the cross section with the polarizations of all particles is
\begin{equation}
\frac{d\sigma}{dy\,d\varphi} = \frac{\alpha^2}{m^2 x^4 y^2}
\sum_{ii'jj'} F_{ij}^{i'j'} \zeta_i \xi_j \zeta'_{i'} \xi'_{j'}\,,
\label{sigres}
\end{equation}
where
\begin{equation}
F_{ij}^{i'j'} = \sum_{kk'} \delta_j \delta_{k'}\;
\frac{1}{2} \Tr \sigma_{j'} \sigma_k \sigma_j \sigma_{k'}
\frac{1}{4} \Tr \rho'_{i'} Q_k \rho_i Q_{k'}
\label{F}
\end{equation}
(the right-hand side depends on $\varphi$ because the polarizations are defined
relative to the scattering plane).
The cross section summed over the final particles' polarizations is
\begin{equation}
\frac{d\sigma}{dy\,d\varphi} = \frac{\alpha^2}{m^2 x^4 y^2} F\,,\quad
F = \sum_{ij} F_{ij}^{00} \zeta_i \xi_j\,.
\label{sigsum}
\end{equation}
The final particles' polarizations are
\begin{equation}
\zeta'_{i'} = \frac{1}{F} \sum_{ij} F_{ij}^{i'0} \zeta_i \xi_j\,,\quad
\xi'_{j'} = \frac{1}{F} \sum_{ij} F_{ij}^{0j'} \zeta_i \xi_j\,.
\label{finpol}
\end{equation}
The components $F_{ij}^{i'j'}$ with $i'\ne0$ and $j'\ne0$ describe the
correlation of the final particles' polarizations.

I have calculated $F_{ij}^{i'j'}$ using REDUCE.
It is convenient to express all vectors via the basis~(\ref{n}).%
\footnote{It is natural to introduce a unit spacelike vector $n$
by \texttt{mass n=i; mshell n;}
The current version of REDUCE (3.4.1) returns \texttt{n.n} as \texttt{i}$^2$.
I am grateful to A.~C.~Hearn for sending the patch to fix this bug.}
I used the package RLFI by R.~Liska from the REDUCE library~\cite{Hearn}
to produce the \LaTeX{} source of the result.
All nonzero components $F_{ij}^{i'j'}$ are presented in the Appendix.

I am grateful to I.~F.~Ginzburg, G.~L.~Kotkin, S.~I.~Polityko, and V.~G.~Serbo
for useful discussions.

\section*{Appendix}

\begin{align*}
&F_{00}^{00} = F_{33}^{33} =
x^{3} y-4 x^{2} y+4 x^{2}+x y^{3}+4 x y^{2}-8 x y+4 y^{2}\\
&F_{00}^{03} = F_{03}^{00} = F_{30}^{33} = F_{33}^{30} =
-4 v^{2} w^{2}\\
&F_{00}^{12} = F_{12}^{00} = F_{21}^{33} = F_{33}^{21} =
\left(x y-2 x+2 y\right) \left(x+y\right) v^{2}\\
&F_{00}^{22} = F_{02}^{31} = F_{11}^{33} = -F_{22}^{00} = -F_{31}^{02} = -F_{33}^{11} =
-2 v^{3} w y\\
&F_{01}^{01} = F_{32}^{32} =
2 x y \left(x y-2 x+2 y\right)\\
&F_{01}^{32} = F_{03}^{22} = F_{11}^{30} = -F_{22}^{03} = -F_{30}^{11} = -F_{32}^{01} =
2 v^{3} w x\\
&F_{02}^{02} = F_{31}^{31} =
\left(x^{2}+y^{2}\right) \left(x y-2 x+2 y\right)\\
&F_{02}^{10} = F_{10}^{02} = -F_{23}^{31} = -F_{31}^{23} =
v^{2} \left(x^{3} y+x^{2} y^{2}-4 x^{2} y+4 x^{2}+4 x y^{2}-8 x y+
4 y^{2}\right)/x\\
&F_{02}^{13} = F_{13}^{02} = -F_{20}^{31} = -F_{31}^{20} =
-4 v^{4} w^{2}/x\\
&F_{02}^{20} = F_{02}^{23} = -F_{10}^{31} = -F_{13}^{31} =
-F_{20}^{02} = -F_{23}^{02} = F_{31}^{10} = F_{31}^{13} =
2 v^{3} w \left(-x y+2 x-2 y\right)/x\\
&F_{03}^{03} = F_{30}^{30} =
2 \left(x^{2} y^{2}-2 x^{2} y+2 x^{2}+2 x y^{2}-4 x y+2 y^{2}\right)\\
&F_{10}^{10} = F_{23}^{23} =
\left(x^{4}+x^{2} y^{2}-4 x^{2} y+4 x^{2}+4 x y^{2}-8 x y+4 y^{2}
\right) \left(x y-2 x+2 y\right)/x^{2}\\
&F_{10}^{13} = F_{13}^{10} = F_{20}^{23} = F_{23}^{20} =
4 v^{2} w^{2} \left(-x y+2 x-2 y\right)/x^{2}\\
&F_{10}^{20} = F_{13}^{23} = -F_{20}^{10} = -F_{23}^{13} =
2 v w \left(-x^{3} y-x^{2} y^{2}+4 x^{2} y-4 x^{2}-4 x y^{2}+8 x y
-4 y^{2}\right)/x^{2}\\
&F_{10}^{23} = -F_{23}^{10} =
2 v^{3} w \left(x^{2} y+4 x y-4 x+4 y\right)/x^{2}\displaybreak\\
&F_{11}^{11} = F_{22}^{22} =
2 \left(x^{3} y^{2}-2 x^{3} y+2 x^{3}-2 x^{2} y+2 x y^{3}-2 x y^{2}
+2 y^{3}\right)/x\\
&F_{11}^{21} = F_{12}^{22} = -F_{21}^{11} = -F_{22}^{12} =
-2 \left(x y-2 x+2 y\right) \left(x+y\right) v w/x\\
&F_{12}^{12} = F_{21}^{21} =
\left(x^{4} y-4 x^{3} y+4 x^{3}+x^{2} y^{3}-4 x^{2} y+4 x y^{3}-4 x y^{2}
+4 y^{3}\right)/x\\
&F_{13}^{13} = F_{20}^{20} =
2 \left(x^{3} y-2 x^{2} y+2 x^{2}+2 x y^{2}-4 x y+2 y^{2}\right) 
\left(x y-2 x+2 y\right)/x^{2}\\
&F_{13}^{20} = -F_{20}^{13} =
2 v^{3} w \left(x^{3}+4 x y-4 x+4 y\right)/x^{2}
\end{align*}

\end{document}